\documentstyle[amstex,amsfonts,preprint,aps]{revtex}
\begin{document}
\title{The embedding of the spacetime in five-dimensional spaces with arbitrary
non-degenerate Ricci tensor}
\author{F. Dahia and C. Romero}
\address{Departamento de F\'{i}sica, Universidade Federal da Para\'{i}ba\\
C. Postal 5008, Jo\~{a}o Pessoa, PB\\
58059-970, Brazil}
\maketitle

\begin{abstract}
We discuss and prove a theorem which asserts that any n-dimensional
semi-Riemannian manifold can be locally embedded in a (n+1)-dimensional
space with a non-degenerate Ricci tensor which is equal, up to a local
analytic diffeomorphism, to the Ricci tensor of an arbitrary specified
space. This may be regarded as a further extension of the Campbell-Magaard
theorem. We highlight the significance of embedding theorems of increasing
degrees of generality in the context of higher dimensional spacetimes
theories and illustrate the new theorem by establishing the embedding of a
general class of Ricci-flat spacetimes.
\end{abstract}

\section{Introduction}

Modern physical theories which regard our spacetime as a hypersurface
embedded in a five-dimensional manifold constitute nowadays a branch of
theoretical physics undergoing quite an active research. On the other hand
the idea of an extra fifth dimension is not new and goes back to the works
of Kaluza and Klein carried out around the first quarter of the twentieth
century\cite{kaluza,klein}. Kaluza-Klein's seminal work has inspired
theoretical physicists to generalize their conjecture in the construction of
unified theories of the fundamental interactions of nature. Subsequent
developments which assume that the universe contains extra hidden dimensions
include among others eleven-dimensional supergravity and superstring
theories. \cite{collins,appel}. More recently much attention has been
devoted to the so-called Randall-Sundrum braneworld scenario where the
spacetime is viewed as a four-dimensional hypersurface embedded in a
five-dimensional Einstein space\cite{randall}. Non-compactified approaches
to Kaluza-Klein gravity also makes use of embedding mechanisms and have been
largely discussed in the literature\cite{wesson}.

In a sense one could say that all spacetime embedding theories \cite{pavsic}
assume, implicitly or explicitly, a mathematical framework which must
provide consistency for the postulates and basic principles set forth by
such theories. In this connection it is of interest to know whether the
embedding theorems of differential geometry are properly taken into account
when constructing higher dimensional models. The analysis of the geometrical
structure underlying some modern embedding theories has recently attracted
the interest of some authors\cite
{romero,lidsey1,lidsey2,herdeiro,lidsey3,dahia}. It seems that there is now
a quest for embedding theorems with increasing degrees of generality, i.e.,
theorems ensuring that arbitrary n-dimensional spacetimes can be embedable
in classes of (n+1)-dimensional spaces the most general as possible.

Two theorems of historical importance which have played a significant role
in physical theories of higher dimensions should be mentioned. The first is
the well-known Janet-Cartan theorem, which asserts that if the embedding
space is flat, the minimum number of extra dimensions needed{\large \ }to
analytically embed a n-dimensional Riemannian manifold is $d$, with $0\leq
d\leq n\left( n-1\right) /2.$ \cite{spivak}.

The second is a little known but powerful theorem due to Campbell\cite
{campbell} the proof of which was outlined by Campbell and completed by
Magaard\cite{magaard}. The content of the Campbell-Magaard theorem is that
any n-dimensional Riemannian manifold with analytic metric, locally, can be
isometrically embedded into a certain (n+1)-dimensional Ricci-flat manifold.
It is interesting to note that both theorems specify a geometry property to
be satisfied by the embedding space by imposing the restrictions $R_{\mu \nu
\lambda \rho }=0$ in one case and $R_{\mu \nu }=0$ in the other. It is also
worth of noting that by relaxing the flatness condition, assumed in the
Janet-Cartan theorem, and replacing it by the weaker Ricci-flatness
condition, the Campbell-Magaard theorem drastically reduces the codimension
of the embedding space to $d=1.$ This seems to give support to the
mathematical consistency of theories in which the dynamics of the embedding
space is governed by the vacuum Einstein field equations\cite{wesson}.
However, the view adopted by Randall-Sundrum braneworld model\cite{randall}
that the embedding space, i.e. the bulk, should correspond to an Einstein
space sourced by a negative cosmological constant has naturally raised the
question of whether Campbell-Magaard theorem could be extended to include
embeddings in arbitrary Einstein spaces. This conjecture was shown to be, in
fact, a theorem the proof of which is given in ref. \cite{dahia}. Embeddings
into spaces sourced by scalar fields also have been considered and a
different extension of the Campbell-Magaard theorem has been proved\cite
{lidsey3,dahia2}. In seeking higher levels of generalization one is led to
consider the more general situation of embedding spaces whose Ricci tensor
is arbitrary. In this paper we shall be concerned with this problem. In
section II we state and prove a theorem which considers embedding spaces
with arbitrary non-degenerate Ricci tensor, and, in a way, would represent a
further generalization of Campbell-Magaard's result. In section III we
illustrate the theorem by establishing the embedding of a general class of
Ricci-flat spacetimes in a given collection of five-dimensional spaces whose
Ricci tensor is equivalent to a specified non-degenerate and non-constant
Ricci tensor.

We believe that insofar as five-dimensional embedding theories are metric
theories it appears to be of relevance to allow the embedding spaces to have
different geometrical properties, which must ultimately be determined by the
dynamics of the theory in question. Therefore generalizations of the known
embedding theorems might be helpful in building new higher dimensional
models.

\section{Extension of Campbell-Magaard theorem: embedding spaces with
arbitrary non-degenerate Ricci tensor}

In this section we want to investigate the existence of a local analytic
embedding of a n-dimensional semi-Riemannian manifold $\left( M^{n},g\right) 
$ into a class of $\left( n+1\right) $-dimensional spaces whose Ricci tensor
is {\it equivalent }to the Ricci tensor of a $\left( n+1\right) $%
-dimensional space arbitrarily specified.

{\bf Definition}. {\it Consider a }$\left( n+1\right) -${\it dimensional
semi-Riemannian space }$\left( \tilde{M}_{0}^{n+1},\tilde{g}_{0}\right) $%
{\it \ and let }$S_{\alpha \beta }${\it \ denote the components of the Ricci
tensor in a coordinate system }$\left\{ x^{\prime \alpha }\right\} ${\it .
Let }$\left( \tilde{M}^{n+1},\tilde{g}\right) ${\it \ be another }$\left(
n+1\right) -${\it dimensional semi-Riemannian space with }$\tilde{R}_{\alpha
\beta }${\it \ denoting the components of the Ricci-tensor in a coordinate
system }$\left\{ x^{\alpha }\right\} ${\it \ which covers a neighborhood of
a point }$p\in \tilde{M}^{n+1}${\it \ whose coordinates are }$%
x_{p}^{1}=...=x_{p}^{n+1}=0.${\it \ Then, we shall say that }$S_{\alpha
\beta }${\it \ and }$\tilde{R}_{\alpha \beta }${\it \ are equivalent if
there exists an analytic local diffeomorphism }$\overline{f}:\tilde{M}%
_{0}^{n+1}\rightarrow \tilde{M}^{n+1}${\it \ at }$p$ {\it such that } 
\begin{equation}
\tilde{R}_{\alpha \beta }\left( x^{\gamma }\right) =\frac{\partial \overline{%
f}^{\mu }}{\partial x^{\alpha }}\frac{\partial \overline{f}^{\nu }}{\partial
x^{\beta }}S_{\mu \nu }\left( x^{\prime \kappa }\right) ,  \label{equiv}
\end{equation}
{\it where }$x^{\prime \kappa }=\overline{f}^{\kappa }\left( x^{\lambda
}\right) .${\it \ In others words, }$S_{\alpha \beta }${\it \ and }$\tilde{R}%
_{\alpha \beta }${\it \ are said to be equivalent if there exists a analytic
function }$\overline{f}^{\mu }=\overline{f}^{\mu }\left( x^{\alpha }\right) $%
{\it \ such that: i) }$\left| \frac{\partial \overline{f}^{\mu }}{\partial
x^{\alpha }}\right| \neq 0${\it \ at }$0\in ${\it \ }${\Bbb R}^{n+1};${\it \
ii) the condition (\ref{equiv}) holds in a neighborhood of }$0\in ${\it \ }$%
{\Bbb R}^{n+1}$.{\it \ In this case, }$\left( \tilde{M}_{0}^{n+1},\tilde{g}%
_{0}\right) $ {\it and} $\left( \tilde{M}^{n+1},\tilde{g}\right) $ {\it are
said to be ``Ricci-equivalent'' spaces. }\footnote{%
Henceforth we shall follow the convention adopted in ref \cite{dahia} where
Latin and Greek indices run from $0$ to $n$ and $n+1,$ respectively.}

Clearly, from the above, the collection ${\cal M}_{\tilde{g}_{0}}^{n+1}$ of
all spaces which are Ricci-equivalent to a given space $\left( \tilde{M}%
_{0}^{n+1},\tilde{g}_{0}\right) $ is well defined. Therefore it makes sense
to discuss the existence of the embedding of a given arbitrary $n-$%
dimensional semi-Riemannian manifold $\left( \tilde{M}^{n},\tilde{g}\right) $
into the class ${\cal M}_{\tilde{g}_{0}}^{n+1}$. In what follows we shall
show that if the Ricci tensor of $\left( \tilde{M}_{0}^{n+1},\tilde{g}%
_{0}\right) $ is non-degenerate, i.e., the matrix formed by its components
has inverse, then the existence of the embedding can be ensured.

We should note, however, that (\ref{equiv}) defines a notion of equivalence
between the covariant Ricci tensor $S_{\alpha \beta }$ and $\tilde{R}%
_{\alpha \beta }.$ This equivalence does not imply that the contravariant
Ricci tensor $S^{\alpha \beta }$ and $\tilde{R}^{\alpha \beta }$ are also
equivalent. In general they are not, unless the diffeomorphism is an
isometry, a condition which is more restrictive than (\ref{equiv}).

Let us consider $\left( \tilde{M}^{n+1},\tilde{g}\right) $ and choose a
coordinate system in which the metric has the form 
\begin{equation}
ds^{2}=\bar{g}_{ik}dx^{i}dx^{k}+\varepsilon \bar{\phi}^{2}dy^{2},
\end{equation}
where $\varepsilon =\pm 1.$ In these coordinates (\ref{equiv}) may be
written in the following equivalent form: 
\begin{eqnarray}
\!\tilde{R}_{ik}\! &=&\!\bar{R}_{ik}+\varepsilon \bar{g}^{jm}\left( \bar{%
\Omega}_{ik}\bar{\Omega}_{jm}-2\bar{\Omega}_{jk}\bar{\Omega}_{im}\right) -%
\frac{\varepsilon }{\bar{\phi}}\frac{\partial \bar{\Omega}_{ik}}{\partial y}+%
\frac{1}{\bar{\phi}}\bar{\nabla}_{i}\bar{\nabla}_{k}\bar{\phi}=\frac{%
\partial \overline{f}^{\mu }}{\partial x^{i}}\frac{\partial \overline{f}%
^{\nu }}{\partial x^{k}}S_{\mu \nu }\left( \overline{f}^{\alpha }\right) 
\nonumber \\
&&  \label{4Rik} \\
\tilde{R}_{i}^{y} &=&\frac{\varepsilon }{\bar{\phi}}\bar{g}^{jk}\left( \bar{%
\nabla}_{j}\bar{\Omega}_{ik}-\bar{\nabla}_{i}\bar{\Omega}_{jk}\right) =\frac{%
\varepsilon }{\bar{\phi}^{2}}\frac{\partial \overline{f}^{\mu }}{\partial y}%
\frac{\partial \overline{f}^{\nu }}{\partial x^{i}}S_{\mu \nu }\left( 
\overline{f}^{\alpha }\right)  \label{4Rin+1} \\
\tilde{G}_{y}^{y} &=&-\frac{1}{2}\bar{g}^{ik}\bar{g}^{jm}\left( \bar{R}%
_{ijkm}+\varepsilon \left( \bar{\Omega}_{ik}\bar{\Omega}_{jm}-\bar{\Omega}%
_{jk}\bar{\Omega}_{im}\right) \right) =\frac{1}{2}\frac{\varepsilon }{\bar{%
\phi}^{2}}\frac{\partial \overline{f}^{\mu }}{\partial y}\frac{\partial 
\overline{f}^{\nu }}{\partial y}S_{\mu \nu }\left( \overline{f}^{\alpha
}\right)  \nonumber \\
&&\hspace{0in}\hspace{5cm}-\frac{1}{2}\bar{g}^{jm}\frac{\partial \overline{f}%
^{\mu }}{\partial x^{j}}\frac{\partial \overline{f}^{\nu }}{\partial x^{m}}%
S_{\mu \nu }\left( \overline{f}^{\alpha }\right) ,  \label{4Gn+1}
\end{eqnarray}
where 
\begin{equation}
\bar{\Omega}_{ik}=-\frac{1}{2\bar{\phi}}\frac{\partial \bar{g}_{ik}}{%
\partial y},
\end{equation}
$G_{\alpha \beta }$ is the Einstein tensor and a bar is used to denote all
the geometrical quantities calculated with the induced metric $\bar{g}_{ik}$
on a generic hypersurface $\Sigma _{c}$ of the foliation $y=c=const.$ Before
we state the main theorem we need a few preliminaries.

We begin by defining the tensor 
\begin{equation}
\tilde{F}_{\beta }^{\alpha }=\tilde{G}_{\beta }^{\alpha }-\left( \tilde{g}%
^{\alpha \gamma }\frac{\partial \overline{f}^{\mu }}{\partial x^{\gamma }}%
\frac{\partial \overline{f}^{\nu }}{\partial x^{\beta }}S_{\mu \nu }-\frac{1%
}{2}\delta _{\beta }^{\alpha }\tilde{g}^{\gamma \lambda }\frac{\partial 
\overline{f}^{\mu }}{\partial x^{\gamma }}\frac{\partial \overline{f}^{\nu }%
}{\partial x^{\lambda }}S_{\mu \nu }\right) .  \label{4F}
\end{equation}
If we now impose that the functions $\overline{f}^{\alpha }$satisfy the
equation 
\begin{equation}
\tilde{\nabla}_{\alpha }\left( \tilde{g}^{\alpha \gamma }\frac{\partial 
\overline{f}^{\mu }}{\partial x^{\gamma }}\frac{\partial \overline{f}^{\nu }%
}{\partial x^{\beta }}S_{\mu \nu }-\frac{1}{2}\delta _{\beta }^{\alpha }%
\tilde{g}^{\gamma \lambda }\frac{\partial \overline{f}^{\mu }}{\partial
x^{\gamma }}\frac{\partial \overline{f}^{\nu }}{\partial x^{\lambda }}S_{\mu
\nu }\right) =0,  \label{difeo}
\end{equation}
then it is easily seen that, as the Einstein tensor $G_{\beta }^{\alpha }$
has vanishing divergence, the tensor $\tilde{F}_{\beta }^{\alpha }$ also is
divergenceless for any metric $\tilde{g}_{\alpha \beta }$, even those which
are not solutions of Eq. (\ref{equiv}). Thus, we are ready to state the
following lemma.

\bigskip {\bf Lemma 1.} {\it Let the functions }$\bar{g}_{ik}\left(
x^{1},...,x^{n},y\right) ${\it , }$\bar{\phi}\left( x^{1},...,x^{n},y\right) 
${\it \ and }$\overline{f}^{\alpha }\left( x^{1},...,x^{n},y\right) $ {\it %
be analytical at }$\left( 0,...,0\right) \in \Sigma _{0}\subset {\Bbb R}%
^{n+1}$.{\it \ Assume that the following conditions hold }

{\it i)} $\bar{g}_{ik}=\bar{g}_{ki};$

{\it ii)} $\det \left( \bar{g}_{ik}\right) \neq 0;$

{\it iii)} $\bar{\phi}\neq 0${\it .}

{\it Suppose further that }$\bar{g}_{ik}$ {\it and }$\overline{f}^{\alpha }$ 
{\it satisfy the equations (\ref{4Rik}) and (\ref{difeo}) in an open set }$%
V\subset {\Bbb R}^{n+1}${\it \ which contains }$0\in {\Bbb R}^{n+1}$, {\it %
and (\ref{4Rin+1}) and (\ref{4Gn+1}) hold at }$\Sigma _{0}.${\it \ Then, }$%
\bar{g}_{ik}${\it , }$\bar{\phi}$ {\it and }$\overline{f}^{\alpha }$ {\it %
satisfy (\ref{4Rin+1}) and (\ref{4Gn+1}) in a neighborhood }$0\in ${\it \ }$%
{\Bbb R}^{n+1}.$

{\it Proof. }The key point of the proof is given by the equation $\tilde{%
\nabla}_{\alpha }\tilde{F}_{\beta }^{\alpha }=0,$ which can be written as 
\begin{equation}
\frac{\partial \tilde{F}_{\beta }^{y}}{\partial y}=-\frac{\partial \tilde{F}%
_{\beta }^{i}}{\partial x^{i}}-\tilde{\Gamma}_{\mu \lambda }^{\mu }\tilde{F}%
_{\beta }^{\lambda }+\tilde{\Gamma}_{\lambda \beta }^{\mu }\tilde{F}_{\mu
}^{\lambda }.
\end{equation}
On the other hand, by assumption (\ref{4Rik}) holds in $V\subset $ ${\Bbb R}%
^{n+1}.$ Then, it can be shown that in $V$ we have $\tilde{F}%
_{k}^{i}=-\delta _{k}^{i}\tilde{F}_{y}^{y}.$ After some algebra we can
deduce that 
\begin{eqnarray}
\frac{\partial \tilde{F}_{y}^{y}}{\partial y} &=&-\varepsilon \bar{\phi}^{2}%
\bar{g}^{ij}\frac{\partial \tilde{F}_{i}^{y}}{\partial x^{j}}-2\tilde{\Gamma}%
_{iy}^{i}\tilde{F}_{y}^{y}+\left( -\varepsilon \frac{\partial \left( \bar{%
\phi}^{2}\bar{g}^{ij}\right) }{\partial y^{j}}-\varepsilon \bar{\phi}^{2}%
\bar{g}^{ij}\tilde{\Gamma}_{kj}^{k}+\tilde{\Gamma}_{yy}^{i}\right) \tilde{F}%
_{i}^{y} \\
\frac{\partial \tilde{F}_{i}^{y}}{\partial y} &=&\frac{\partial \tilde{F}%
_{y}^{y}}{\partial x^{i}}+2\tilde{\Gamma}_{yi}^{y}\tilde{F}_{y}^{y}+\left( 
\tilde{\Gamma}_{yi}^{k}+\varepsilon \bar{\phi}^{2}\bar{g}^{kj}\tilde{\Gamma}%
_{ij}^{y}-\tilde{\Gamma}_{y\mu }^{\mu }\delta _{i}^{k}\right) \tilde{F}%
_{k}^{y}.
\end{eqnarray}
Since at the hypersurface $\Sigma _{0}$ the equations (\ref{4Rin+1}) and (%
\ref{4Gn+1}) also hold, it follows that $\tilde{F}_{\beta }^{y}=0$ at $%
\Sigma _{0}$ and hence $\left. \frac{\partial \tilde{F}_{\beta }^{y}}{%
\partial y}\right| _{y=0}=0.$ It is not difficult to show by mathematical
induction that all the derivatives (to any order) of $\tilde{F}_{\beta }^{y}$
vanish at $y=0.$ As $\tilde{F}_{\beta }^{y}$ is analytic we conclude that $%
\tilde{F}_{\beta }^{y}=0$ in an open set of ${\Bbb R}^{n+1}.$ Hence, Eqs. (%
\ref{4Rin+1}) and (\ref{4Gn+1}), which are equivalent to $\tilde{F}_{\beta
}^{y}=0,$ also hold in an open set of ${\Bbb R}^{n+1}$ which includes the
origin. This proves the lemma.

The question which now arises is: do Eqs. (\ref{4Rik}) and (\ref{difeo})
admit solution? To answer this question we first note that (\ref{4Rik}) can
be expressed in the following form 
\begin{eqnarray}
\frac{\partial ^{2}\bar{g}_{ik}}{\partial y^{2}} &=&-2\varepsilon \bar{\phi}%
^{2}\left( \frac{\partial \overline{f}^{\mu }}{\partial x^{i}}\frac{\partial 
\overline{f}^{\nu }}{\partial x^{k}}S_{\mu \nu }\left( \overline{f}^{\alpha
}\right) \right) +\frac{1}{\bar{\phi}}\frac{\partial \bar{\phi}}{\partial y}%
\frac{\partial \bar{g}_{ik}}{\partial y}-\frac{1}{2}\bar{g}^{jm}\left( \frac{%
\partial \bar{g}_{ik}}{\partial y}\frac{\partial \bar{g}_{jm}}{\partial y}-2%
\frac{\partial \bar{g}_{im}}{\partial y}\frac{\partial \bar{g}_{jk}}{%
\partial y}\right)  \nonumber \\
&&-2\varepsilon \bar{\phi}\left( \frac{\partial ^{2}\bar{\phi}}{\partial
x^{i}\partial x^{k}}-\frac{\partial \bar{\phi}}{\partial x^{j}}\bar{\Gamma}%
_{ik}^{j}\right) -2\varepsilon \bar{\phi}^{2}\bar{R}_{ik}.  \label{gik-can}
\end{eqnarray}

Second, let us rewrite Eq. (\ref{difeo}) in the form 
\begin{eqnarray}
\frac{\partial }{\partial x^{\alpha }}\left( \tilde{g}^{\alpha \gamma }\frac{%
\partial \overline{f}^{\mu }}{\partial x^{\gamma }}\frac{\partial \overline{f%
}^{\nu }}{\partial x^{\beta }}S_{\mu \nu }-\frac{1}{2}\delta _{\beta
}^{\alpha }\tilde{g}^{\gamma \lambda }\frac{\partial \overline{f}^{\mu }}{%
\partial x^{\gamma }}\frac{\partial \overline{f}^{\nu }}{\partial x^{\lambda
}}S_{\mu \nu }\right) &&+  \nonumber \\
+\tilde{\Gamma}_{\alpha _{\sigma }}^{\alpha }\left( \tilde{g}^{\sigma \gamma
}\frac{\partial \overline{f}^{\mu }}{\partial x^{\gamma }}\frac{\partial 
\overline{f}^{\nu }}{\partial x^{\beta }}S_{\mu \nu }\right) -\tilde{\Gamma}%
_{\alpha _{\beta }}^{\sigma }\left( \tilde{g}^{\alpha \gamma }\frac{\partial 
\overline{f}^{\mu }}{\partial x^{\gamma }}\frac{\partial \overline{f}^{\nu }%
}{\partial x^{\sigma }}S_{\mu \nu }\right) &=&0.
\end{eqnarray}
We now isolate the terms which contain second-order derivatives of $%
\overline{f}^{\alpha }$ with respect to $y$ in the equation above. Putting $%
\beta =n+1$ we obtain 
\begin{eqnarray}
\frac{\varepsilon }{\bar{\phi}^{2}}\frac{\partial ^{2}\overline{f}^{\mu }}{%
\partial y^{2}}\frac{\partial \overline{f}^{\nu }}{\partial y}S_{\mu \nu }
&=&-\frac{1}{2}\frac{\partial \overline{f}^{\mu }}{\partial y}\frac{\partial 
\overline{f}^{\nu }}{\partial y}\frac{\partial }{\partial y}\left( \frac{%
\varepsilon }{\bar{\phi}^{2}}S_{\mu \nu }\right) +\frac{1}{2}\frac{\partial 
}{\partial y}\left( \bar{g}^{jk}\frac{\partial \overline{f}^{\mu }}{\partial
x^{j}}\frac{\partial \overline{f}^{\nu }}{\partial x^{k}}S_{\mu \nu }\right)
-\frac{\partial }{\partial x^{j}}\left( \bar{g}^{jk}\frac{\partial \overline{%
f}^{\mu }}{\partial x^{k}}\frac{\partial \overline{f}^{\nu }}{\partial y}%
S_{\mu \nu }\right)  \nonumber \\
&&-\tilde{\Gamma}_{\alpha _{\sigma }}^{\alpha }\left( \tilde{g}^{\sigma
\gamma }\frac{\partial \overline{f}^{\mu }}{\partial x^{\gamma }}\frac{%
\partial \overline{f}^{\nu }}{\partial x^{\beta }}S_{\mu \nu }\right) +%
\tilde{\Gamma}_{\alpha _{\beta }}^{\sigma }\left( \tilde{g}^{\alpha \gamma }%
\frac{\partial \overline{f}^{\mu }}{\partial x^{\gamma }}\frac{\partial 
\overline{f}^{\nu }}{\partial x^{\sigma }}S_{\mu \nu }\right) .
\label{difeo-n+1}
\end{eqnarray}
For $\beta =i$ we have 
\begin{eqnarray}
\frac{\varepsilon }{\bar{\phi}^{2}}\frac{\partial ^{2}\overline{f}^{\mu }}{%
\partial y^{2}}\frac{\partial \overline{f}^{\nu }}{\partial x^{i}}S_{\mu \nu
} &=&-\frac{\partial \overline{f}^{\mu }}{\partial y}\frac{\partial }{%
\partial y}\left( \frac{\varepsilon }{\bar{\phi}^{2}}\frac{\partial 
\overline{f}^{\nu }}{\partial x^{i}}S_{\mu \nu }\right) -\frac{\partial }{%
\partial x^{j}}\left( \bar{g}^{jk}\frac{\partial \overline{f}^{\mu }}{%
\partial x^{j}}\frac{\partial \overline{f}^{\nu }}{\partial x^{i}}S_{\mu \nu
}\right) +\frac{1}{2}\frac{\partial }{\partial x^{i}}\left( \tilde{g}%
^{\sigma \gamma }\frac{\partial \overline{f}^{\mu }}{\partial x^{\gamma }}%
\frac{\partial \overline{f}^{\nu }}{\partial x^{\beta }}S_{\mu \nu }\right) 
\nonumber \\
&&-\tilde{\Gamma}_{\alpha _{\sigma }}^{\alpha }\left( \tilde{g}^{\sigma
\gamma }\frac{\partial \overline{f}^{\mu }}{\partial x^{\gamma }}\frac{%
\partial \overline{f}^{\nu }}{\partial x^{\beta }}S_{\mu \nu }\right) +%
\tilde{\Gamma}_{\alpha _{\beta }}^{\sigma }\left( \tilde{g}^{\alpha \gamma }%
\frac{\partial \overline{f}^{\mu }}{\partial x^{\gamma }}\frac{\partial 
\overline{f}^{\nu }}{\partial x^{\sigma }}S_{\mu \nu }\right) .
\label{difeo-i}
\end{eqnarray}

Clearly, the right-hand side of (\ref{difeo-n+1}) and (\ref{difeo-i}) does
not contain second-order derivatives of the functions $\overline{f}^{\alpha
} $ and $\bar{g}_{ik}$ with respect to $y$. Therefore, they are of the form 
\begin{equation}
\frac{\partial ^{2}\overline{f}^{\mu }}{\partial y^{2}}\frac{\partial 
\overline{f}^{\nu }}{\partial x^{\beta }}S_{\mu \nu }=Q_{\beta }\left( 
\overline{f}^{\lambda },\frac{\partial \overline{f}^{\lambda }}{\partial
x^{\sigma }},\frac{\partial ^{2}\overline{f}^{\lambda }}{\partial x^{\sigma
}\partial x^{i}},\bar{g}_{ik},\frac{\partial \bar{g}_{ik}}{\partial
x^{\sigma }}\right) .
\end{equation}
(Of course $Q_{\beta }$ also depends on $\bar{\phi}$ and its derivatives,
however this fact is not relevant for our present reasoning). Thus, assuming
that $\left| \frac{\partial \overline{f}^{\mu }}{\partial x^{\alpha }}%
\right| \neq 0$ (we shall see later on that this assumption can always be
made) we can write 
\begin{equation}
\frac{\partial ^{2}\overline{f}^{\mu }}{\partial y^{2}}S_{\mu \nu }=\frac{%
\partial x^{\beta }}{\partial \overline{f}^{\nu }}Q_{\beta }\left( \overline{%
f}^{\lambda },\frac{\partial \overline{f}^{\lambda }}{\partial x^{\sigma }},%
\frac{\partial ^{2}\overline{f}^{\lambda }}{\partial x^{\sigma }\partial
x^{i}},\bar{g}_{ik},\frac{\partial \bar{g}_{ik}}{\partial x^{\sigma }}%
\right) .  \label{difeo-b}
\end{equation}
If we suppose that $S_{\mu \nu }$ is invertible, i.e., there exists $\left(
S^{-1}\right) ^{\nu \lambda }$ such that 
\begin{equation}
S_{\mu \nu }\left( S^{-1}\right) ^{\nu \lambda }=\delta _{\mu }^{\lambda },
\end{equation}
then (\ref{difeo}) can be put into the canonical form 
\begin{equation}
\frac{\partial ^{2}\overline{f}^{\mu }}{\partial y^{2}}=P^{\mu }\left( 
\overline{f}^{\lambda },\frac{\partial \overline{f}^{\lambda }}{\partial
x^{\sigma }},\frac{\partial ^{2}\overline{f}^{\lambda }}{\partial x^{\sigma
}\partial x^{i}},\bar{g}_{ik},\frac{\partial \bar{g}_{ik}}{\partial
x^{\sigma }}\right) ,  \label{difeo-can}
\end{equation}
where each $P^{\mu }$ is analytic with respect to its arguments provided
that $\left| \frac{\partial \overline{f}^{\mu }}{\partial x^{\alpha }}%
\right| \neq 0$, $\left| \bar{g}_{ik}\right| \neq 0$ and $\bar{\phi}\neq 0$.

It easy to see that the Cauchy-Kowalewski theorem (see Appendix) can be
applied to the equations (\ref{difeo-can}) and (\ref{gik-can}), which are
equivalent to (\ref{4Rik}) and (\ref{difeo}), respectively. According to the
above-mentioned theorem, if an analytic function $\bar{\phi}\neq 0$ is
chosen, then there exists a unique set of analytic functions $\bar{g}_{ik}$
and $\overline{f}^{\alpha }$ that are solutions of (\ref{4Rik}) and (\ref
{difeo}) satisfying the initial conditions 
\begin{eqnarray}
\bar{g}_{ik}\left( x^{1},..,x^{n},0\right) &=&g_{ik}\left(
x^{1},..,x^{n}\right)  \label{4cig} \\
\frac{\partial \bar{g}_{ik}}{\partial y}\left( x^{1},..,x^{n},0\right) &=&-2%
\bar{\phi}\left( x^{1},..,x^{n},0\right) \Omega _{ik}\left(
x^{1},...,x^{n}\right)  \label{4cih} \\
\overline{f}^{\alpha }\left( x^{1},..,x^{n},0\right) &=&\xi ^{\alpha }\left(
x^{1},..,x^{n}\right)  \label{4ciphi} \\
\frac{\partial \overline{f}^{\alpha }}{\partial y}\left(
x^{1},..,x^{n},0\right) &=&\eta ^{\alpha }\left( x^{1},..,x^{n}\right) ,
\label{4cidphi}
\end{eqnarray}
where $g_{ik},$ $\Omega _{ik},$ $\xi ^{\alpha }$ and $\eta ^{\alpha }$ are
analytic functions at the origin $0\in {\Bbb R}^{n},$ and the following
conditions hold: i)$\left| \frac{\partial \overline{f}^{\mu }}{\partial
x^{\alpha }}\right| _{0}\neq 0$; e ii) $\left| g_{ik}\right| \neq 0.$
(Incidentally, we can easily verify that the condition (i) is satisfied by
simply choosing: $\xi ^{i}=x^{i};$ $\xi ^{n+1}=0;$ $\eta ^{i}=0$ and $\eta
^{n+1}=1.$ With this choice, $\left| \frac{\partial \overline{f}^{\mu }}{%
\partial x^{\alpha }}\right| _{0}=1.$)

We now are ready to state the following theorem.

{\bf Theorem 1 }{\it Let }$M^{n}$ {\it be a n-dimensional semi-Riemannian
manifold with metric given by} 
\[
ds^{2}=g_{ik}dx^{i}dx^{k}, 
\]
{\it \ in a coordinate system }$\left\{ x^{i}\right\} $ {\it of }$M^{n}.$ 
{\it Let }$p\in M^{n},$ {\it have coordinates }$x_{p}^{1}=...=x_{p}^{n}=0.$ 
{\it Then, }$M^{n}$ {\it has a local isometric and analytic embedding (at
the point }$p$) {\it in a (n+1)-dimensional space }$\left( \tilde{M}^{n+1},%
\tilde{g}\right) ${\it \ whose Ricci tensor is equivalent to the symmetric,
analytic and non-degenerate tensor }$S_{\mu \nu }$ {\it if and only if there
exist functions }$\Omega _{ik}\left( x^{1},...,x^{n}\right) $ {\it \ }$%
(i,k=1,..,n),$ $\xi ^{\alpha }\left( x^{1},..,x^{n}\right) $, $\eta ^{\alpha
}\left( x^{1},..,x^{n}\right) $ $(\alpha =1,...,n+1)$ {\it and} $\phi \left(
x^{1},..,x^{n}\right) \neq 0$ {\it that are analytic at }$0\in {\Bbb R}^{n},$%
{\it \ such that } 
\begin{eqnarray}
\Omega _{ik} &=&\Omega _{ki}  \label{4eq1} \\
g^{jk}\left( \nabla _{j}\Omega _{ik}-\nabla _{i}\Omega _{jk}\right) &=&\frac{%
1}{\phi }\eta ^{\mu }\frac{\partial \xi ^{\nu }}{\partial x^{i}}S_{\mu \nu
}\left( \xi ^{\alpha }\right)  \label{4eq2} \\
g^{ik}g^{jm}\left( R_{ijkm}+\varepsilon \left( \Omega _{ik}\Omega
_{jm}-\Omega _{jk}\Omega _{im}\right) \right) &=&-\frac{\varepsilon }{\phi
^{2}}\eta ^{\mu }\eta ^{v}S_{\mu \nu }\left( \xi ^{\alpha }\right) +g^{jm}%
\frac{\partial \xi ^{\mu }}{\partial x^{j}}\frac{\partial \xi ^{\nu }}{%
\partial x^{m}}S_{\mu \nu }\left( \xi ^{\alpha }\right)  \nonumber \\
&&  \label{4eq3} \\
\left| 
\begin{array}{ccc}
\frac{\partial \xi ^{1}}{\partial x^{1}} & \cdots & \frac{\partial \xi ^{n+1}%
}{\partial x^{1}} \\ 
\vdots &  & \vdots \\ 
\frac{\partial \xi ^{1}}{\partial x^{n}} & \cdots & \frac{\partial \xi ^{n+1}%
}{\partial x^{n}} \\ 
\eta ^{1} & \cdots & \eta ^{n+1}
\end{array}
\right| &\neq &0  \label{4eq4}
\end{eqnarray}

{\it Proof.} Let us start with the necessary condition. If $\left(
M^{n},g\right) $ has an embedding in $\left( \tilde{M}^{n+1},\tilde{g}%
\right) ,$ then it can be proved that there exists a coordinate system in
which the metric of the embedding space has the form\cite{dahia} 
\begin{equation}
ds^{2}=\bar{g}_{ik}dx^{i}dx^{k}+\varepsilon \bar{\phi}^{2}dy^{2},
\label{4ds2}
\end{equation}
where the analytic functions $\bar{g}_{ik}\left( x^{1},...,x^{n},y\right) $
and $\bar{\phi}\left( x^{1},...,x^{n},y\right) $ are such that $\bar{\phi}%
\left( x^{1},...,x^{n},y\right) \neq 0$ and that $\bar{g}_{ik}\left(
x^{1},...,x^{n},0\right) =g_{ik}\left( x^{1},...,x^{n}\right) $ in an open
set of ${\Bbb R}^{n}$ which contains the origin. Given that the Ricci tensor
of the embedding space $\left( \tilde{M}^{n+1},\tilde{g}\right) $ is, by
assumption, equivalent to $S_{\mu \nu },$ then the equations (\ref{4Rik}), (%
\ref{4Rin+1}), (\ref{4Gn+1}) and (\ref{difeo}) are satisfied in a
neighborhood of $0\in {\Bbb R}^{n+1}$ for some functions $\overline{f}^{\mu
}.$ In particular, the equations (\ref{4Rin+1}) and (\ref{4Gn+1}) hold for $%
y=0.$ Therefore, if we define $\Omega _{ik},$ $\xi ^{\alpha }$ , $\eta
^{\alpha }$ by the relations (\ref{4cih}), (\ref{4ciphi}) and (\ref{4cidphi}%
), and take $\phi \left( x^{1},..,x^{n}\right) =\bar{\phi}\left(
x^{1},...,x^{n},0\right) $ then the Eqs.(\ref{4eq1}), (\ref{4eq2}), (\ref
{4eq3}) and (\ref{4eq4}) are satisfied.

Let us turn to the sufficiency. Suppose there exist functions $\Omega
_{ik}\left( x^{1},...,x^{n}\right) ,$ $\xi ^{\alpha }\left(
x^{1},..,x^{n}\right) $, $\eta ^{\alpha }\left( x^{1},..,x^{n}\right) $ and $%
\phi \left( x^{1},..,x^{n}\right) \neq 0$ which satisfy (\ref{4eq1}), (\ref
{4eq2}), (\ref{4eq3}) and (\ref{4eq4}). Choose an analytic function $\bar{%
\phi}\left( x^{1},...,x^{n},y\right) \neq 0$ such that $\bar{\phi}\left(
x^{1},...,x^{n},0\right) =\phi \left( x^{1},..,x^{n}\right) .$ By virtue of
the Cauchy-Kowalewski theorem there exists a unique set of analytic
functions $\bar{g}_{ik}\left( x^{1},...,x^{n},y\right) $ and $\overline{f}%
^{\alpha }\left( x^{1},...,x^{n},y\right) $ satisfying the equations (\ref
{4Rik}), (\ref{difeo}) and the initial conditions (\ref{4cig}), (\ref{4cih}%
), (\ref{4ciphi}) and (\ref{4cidphi}). Since, by assumption, the initial
conditions satisfy the equations (\ref{4eq1}), (\ref{4eq2}) and (\ref{4eq3})
then $\bar{g}_{ik},\bar{\phi}$ and $\overline{f}^{\alpha }$ satisfy (\ref
{4Rin+1}) and (\ref{4Gn+1}) at $y=0.$ It follows from lemma 1 that $\bar{g}%
_{ik},\bar{\phi}$ and $\overline{f}^{\alpha }$ satisfy (\ref{equiv}) in an
open set of ${\Bbb R}^{n+1}$ which contains the origin. Further we can say
that $\overline{f}^{\alpha }$ is a diffeomorphism since by virtue (\ref{4eq4}%
) we have $\left| \frac{\partial \overline{f}^{\mu }}{\partial x^{\alpha }}%
\right| \neq 0.$ Therefore, we conclude that the (n+1)-dimensional manifold
whose line element (\ref{4ds2}) is formed with the solutions $\bar{g}_{ik}$
and $\bar{\phi}$ is a space whose Ricci tensor is equivalent to $S_{\mu \nu
} $, and the embedding of the manifold $\left( M^{n},g\right) $ is given by $%
y=0.$ This completes the proof.

We now need to show that once the functions $g_{ik}$ are given the system of
equations (\ref{4eq1}), (\ref{4eq2}), (\ref{4eq3}) and (\ref{4eq4}) always
admits solution for $\Omega _{ik}.$ For simplicity we take $\xi ^{i}=x^{i};$ 
$\xi ^{n+1}=0;$ $\eta ^{i}=0$ e $\eta ^{n+1}=1.$ With this choice the
condition (\ref{4eq4}) is readily satisfied. The equations (\ref{4eq1}), (%
\ref{4eq2}) and (\ref{4eq3}) constitute a set of $n$ partial differential
equations (\ref{4eq2}) plus a constraint equation (\ref{4eq3}) for $n\left(
n+1\right) /2$ independent functions $\Omega _{ik}.$ Except for $n=1,$ the
number of unknown functions is greater than (or equal to (n=2)) the number
of equations. Then, out of the set of functions $\Omega _{ik}$ we pick $n$
functions $\Omega _{1k}$ $\left( k\geq 2\right) $ and $\Omega _{r^{\prime
}n} $ to be regarded as the unknowns\footnote{%
The $r^{\prime }$ index has the following meaning. We assume, for the sake
of the argument, that we are using a coordinate system in which $g_{11}\neq
0 $ and $g_{1k}=0,$ $k=2,...,n.$ Hence, there exists at least an index $%
r^{\prime }>1$ such that $g^{r^{\prime }n}\neq 0$, since $\left|
g_{ik}\right| \neq 0$.}. The next step is to write (\ref{4eq2}) in a
suitable form for application of the Cauchy-Kowalewski theorem (first-order
derivative version) to ensure the existence of the solution. For the sake of
brevity we shall omit the detailed proof and refer the reader to references 
\cite{dahia,magaard} where a similar procedure is carried out. Then it can
be shown that after solving (\ref{4eq2}) for $\Omega _{1k}$ $\left( k\geq
2\right) $ and $\Omega _{r^{\prime }n}$ we obtain 
\begin{align}
\frac{\partial \Omega _{r^{\prime }n}}{\partial x^{1}}& =\frac{1}{%
g^{r^{\prime }n}\left( \delta _{r^{\prime }n}-2\right) }\left[ -%
%TCIMACRO{\underset{r,s>1}{g^{rs}}}%
%BeginExpansion
\mathrel{\mathop{g^{rs}}\limits_{r,s>1}}%
%EndExpansion
\Omega _{1s,r}+2%
%TCIMACRO{
%\underset{\QTATOPD. . {1<r<s}{r,s\neq r^{\prime },n}}{g^{rs}\Omega _{rs,1}}}%
%BeginExpansion
\mathrel{\mathop{g^{rs}\Omega _{rs,1}}\limits_{%
{\textstyle{1<r<s \atopwithdelims.. r,s\neq r^{\prime },n}}}}%
%EndExpansion
+g^{rr}%
%TCIMACRO{\underset{\QTATOPD. . {r>1}{r\neq r^{\prime }}}{\Omega _{rr,1}}}%
%BeginExpansion
\mathrel{\mathop{\Omega _{rr,1}}\limits_{%
{\textstyle{r>1 \atopwithdelims.. r\neq r^{\prime }}}}}%
%EndExpansion
+\right.  \label{481} \\
& \left. +g^{r^{\prime }r^{\prime }}\Omega _{r^{\prime }r^{\prime },1}\left(
1-\delta _{r^{\prime }n}\right) -%
%TCIMACRO{\underset{r,s>1}{g^{rs}}}%
%BeginExpansion
\mathrel{\mathop{g^{rs}}\limits_{r,s>1}}%
%EndExpansion
\left( 
%TCIMACRO{\underset{t\leq r}{\Omega _{tr}}}%
%BeginExpansion
\mathrel{\mathop{\Omega _{tr}}\limits_{t\leq r}}%
%EndExpansion
\Gamma _{s1}^{t}+%
%TCIMACRO{\underset{r<t}{\Omega _{rt}}}%
%BeginExpansion
\mathrel{\mathop{\Omega _{rt}}\limits_{r<t}}%
%EndExpansion
\Gamma _{s1}^{t}-\Omega _{11}\Gamma _{sr}^{1}-%
%TCIMACRO{\underset{t<1}{\Omega _{1t}}}%
%BeginExpansion
\mathrel{\mathop{\Omega _{1t}}\limits_{t<1}}%
%EndExpansion
\Gamma _{sr}^{t}\right) +\frac{1}{\phi }S_{1y}\left( x^{i}\right) \right] 
\nonumber
\end{align}
where no sum over $r^{\prime }$ is implied, and 
\begin{eqnarray}
\frac{\partial \Omega _{1k}}{\partial x^{1}} &=&g_{11}\left[ -%
%TCIMACRO{\underset{r,s>1}{g^{rs}}}%
%BeginExpansion
\mathrel{\mathop{g^{rs}}\limits_{r,s>1}}%
%EndExpansion
\left( 
%TCIMACRO{\underset{s\leq k}{\Omega _{sk,r}}}%
%BeginExpansion
\mathrel{\mathop{\Omega _{sk,r}}\limits_{s\leq k}}%
%EndExpansion
+%
%TCIMACRO{\underset{k<s}{\Omega _{ks,r}}}%
%BeginExpansion
\mathrel{\mathop{\Omega _{ks,r}}\limits_{k<s}}%
%EndExpansion
-2%
%TCIMACRO{\underset{r<s}{\Omega _{rs,k}}}%
%BeginExpansion
\mathrel{\mathop{\Omega _{rs,k}}\limits_{r<s}}%
%EndExpansion
\right) -g^{11}\Omega _{11,k}-g^{rr}%
%TCIMACRO{\underset{r>1}{\Omega _{rr,k}}}%
%BeginExpansion
\mathrel{\mathop{\Omega _{rr,k}}\limits_{r>1}}%
%EndExpansion
\right.  \label{482} \\
&&\left. -g^{rs}\left( 
%TCIMACRO{\underset{t\leq r}{\Omega _{tr}}}%
%BeginExpansion
\mathrel{\mathop{\Omega _{tr}}\limits_{t\leq r}}%
%EndExpansion
\Gamma _{sk}^{t}+%
%TCIMACRO{\underset{r<t}{\Omega _{rt}}}%
%BeginExpansion
\mathrel{\mathop{\Omega _{rt}}\limits_{r<t}}%
%EndExpansion
\Gamma _{sk}^{t}-%
%TCIMACRO{\underset{t\leq k}{\Omega _{tk}}}%
%BeginExpansion
\mathrel{\mathop{\Omega _{tk}}\limits_{t\leq k}}%
%EndExpansion
\Gamma _{sr}^{t}-%
%TCIMACRO{\underset{k<t}{\Omega _{kt}}}%
%BeginExpansion
\mathrel{\mathop{\Omega _{kt}}\limits_{k<t}}%
%EndExpansion
\Gamma _{sr}^{t}\right) +\frac{1}{\phi }S_{k\,y}\left( x^{i}\right) \right]
,\qquad k\geq 2.  \nonumber
\end{eqnarray}
where $\Omega _{11}$ must be substituted by 
\begin{eqnarray}
\Omega _{11} &=&\frac{1}{2g^{11}%
%TCIMACRO{\underset{r,s>1}{g^{rs}}}%
%BeginExpansion
\mathrel{\mathop{g^{rs}}\limits_{r,s>1}}%
%EndExpansion
\left( 
%TCIMACRO{\underset{r\leq s}{\Omega _{rs}}}%
%BeginExpansion
\mathrel{\mathop{\Omega _{rs}}\limits_{r\leq s}}%
%EndExpansion
+%
%TCIMACRO{\underset{s<r}{\Omega _{sr}}}%
%BeginExpansion
\mathrel{\mathop{\Omega _{sr}}\limits_{s<r}}%
%EndExpansion
\right) }\left[ 2g^{11}%
%TCIMACRO{\underset{r,s>1}{g^{rs}}}%
%BeginExpansion
\mathrel{\mathop{g^{rs}}\limits_{r,s>1}}%
%EndExpansion
\Omega _{1r}\Omega _{1s}\right.  \nonumber \\
&&-%
%TCIMACRO{\underset{r,s,t,u>1}{g^{rs}g^{tu}}}%
%BeginExpansion
\mathrel{\mathop{g^{rs}g^{tu}}\limits_{r,s,t,u>1}}%
%EndExpansion
\left[ \left( 
%TCIMACRO{\underset{r\leq s}{\Omega _{rs}}}%
%BeginExpansion
\mathrel{\mathop{\Omega _{rs}}\limits_{r\leq s}}%
%EndExpansion
+%
%TCIMACRO{\underset{s<r}{\Omega _{sr}}}%
%BeginExpansion
\mathrel{\mathop{\Omega _{sr}}\limits_{s<r}}%
%EndExpansion
\right) \left( 
%TCIMACRO{\underset{t\leq u}{\Omega _{tu}}}%
%BeginExpansion
\mathrel{\mathop{\Omega _{tu}}\limits_{t\leq u}}%
%EndExpansion
+%
%TCIMACRO{\underset{u<t}{\Omega _{ut}}}%
%BeginExpansion
\mathrel{\mathop{\Omega _{ut}}\limits_{u<t}}%
%EndExpansion
\right) -\left( 
%TCIMACRO{\underset{r\leq u}{\Omega _{ru}}}%
%BeginExpansion
\mathrel{\mathop{\Omega _{ru}}\limits_{r\leq u}}%
%EndExpansion
+%
%TCIMACRO{\underset{u<r}{\Omega _{ur}}}%
%BeginExpansion
\mathrel{\mathop{\Omega _{ur}}\limits_{u<r}}%
%EndExpansion
\right) \left( 
%TCIMACRO{\underset{s\leq t}{\Omega _{st}}}%
%BeginExpansion
\mathrel{\mathop{\Omega _{st}}\limits_{s\leq t}}%
%EndExpansion
+%
%TCIMACRO{\underset{t<s}{\Omega _{ts}}}%
%BeginExpansion
\mathrel{\mathop{\Omega _{ts}}\limits_{t<s}}%
%EndExpansion
\right) \right]  \nonumber \\
&&\quad \quad \left. 
%TCIMACRO{
%\underset{}{-\varepsilon \left( R-\frac{\varepsilon }{\phi ^{2}}S_{yy}\left( x^{i}\right) +g^{jm}S_{jm}\left( x^{i}\right) \right) }}%
%BeginExpansion
\mathrel{\mathop{-\varepsilon \left( R-\frac{\varepsilon }{\phi ^{2}}S_{yy}\left( x^{i}\right) +g^{jm}S_{jm}\left( x^{i}\right) \right) }\limits_{}}%
%EndExpansion
\right] .  \label{483}
\end{eqnarray}

Finally, if we choose the functions $\Omega _{ik}$ $\left( i\leq
k,i>1,\left( i,k\right) \neq \left( r^{\prime },n\right) \right) ,$ $\phi
\neq 0$ as being analytic at the origin, and since $S_{\mu \nu }\left(
x^{i}\right) $ are also analytic, then in view of the Cauchy-Kowalewski
theorem the system of equations (\ref{481}) and (\ref{482}) admits a
solution that is analytic at the origin. Therefore, given arbitrary analytic
functions $g_{ik}\left( x^{1},...,x^{n}\right) $ the existence of the
functions $\Omega _{ik}\left( x^{1},...,x^{n}\right) $ {\it \ }$%
(i,k=1,..,n), $ $\xi ^{\alpha }\left( x^{1},..,x^{n}\right) $, $\eta
^{\alpha }\left( x^{1},..,x^{n}\right) $ which satisfy (\ref{4eq1}), (\ref
{4eq2}), (\ref{4eq3}) and (\ref{4eq4}) is ensured, so Theorem 1 applies.

It should be mentioned that in the case where $S_{\mu \nu }=0,$ Eq. (\ref
{difeo}) holds for any functions $\overline{f}^{\alpha }\left( x^{\beta
}\right) ;$ hence all the results derived above applies when the space $%
\left( \tilde{M}_{0}^{n+1},\tilde{g}_{0}\right) $ has a vanishing Ricci
tensor. Therefore, we can state the following theorem:

{\bf Theorem 2.} {\it Let }$M^{n}$ {\it be a n-dimensional semi-Riemannian
manifold with metric given by} 
\[
ds^{2}=g_{ik}dx^{i}dx^{k}, 
\]
{\it \ in a coordinate system }$\left\{ x^{i}\right\} $ {\it of }$M^{n}.$ 
{\it Let }$p\in M^{n},$ {\it have coordinates }$x_{p}^{1}=...=x_{p}^{n}=0.$ 
{\it Consider a (n+1)-dimensional semi-Riemannian space }$\left( \tilde{M}%
_{0}^{n+1},\tilde{g}_{0}\right) ${\it \ whose Ricci tensor is either
non-degenerate or null. If }$g_{ik}$ {\it are analytic functions at }$0\in 
{\Bbb R}^{n},$ {\it then }$\left( M^{n},g\right) $ {\it has a local
isometric and analytic embedding (at the point }$p$) {\it in a
(n+1)-dimensional space which is Ricci-equivalent to }$\left( \tilde{M}%
_{0}^{n+1},\tilde{g}_{0}\right) .$

Therefore, we conclude that if the space $\left( \tilde{M}_{0}^{n+1},\tilde{g%
}_{0}\right) $ is a solution of the Einstein equations for some source, then
Theorem 2 guarantees that there exists a space which satisfies the ``same''
Einstein equations up to a coordinate transformation (see Eq. (\ref{equiv})
), in which the spacetime $\left( M^{n},g\right) $ can be embedded

\section{A simple application of Theorem 2}

Up to this point we have considered the Ricci tensor only through its
covariant components $\tilde{R}_{\alpha \beta }.$ However, it is not
difficult to realize that all the previous results we have obtained are
still valid if the mixed $\tilde{R}_{\beta }^{\alpha }$ or contravariant
components $\tilde{R}^{\alpha \beta }$ are considered instead.

In what follows we illustrate Theorem 2 in its Ricci-tensor mixed-components
version.

Consider the five-dimensional semi-Riemannian space $\left( \tilde{M}%
_{0}^{5},\tilde{g}_{0}\right) $ with a metric given by 
\begin{equation}
^{5}ds^{2}=\left( y+1\right) ^{\frac{4}{5}}\left(
-dt^{2}+dx^{2}+dy^{2}+dz^{2}\right) +\frac{24}{25}\varepsilon dy^{2}.
\label{5ds2}
\end{equation}

If we calculate the mixed components of the Ricci $S_{\nu }^{\mu }$ tensor
for this metric we obtain 
\begin{equation}
S_{\nu }^{\mu }=diag\left( \frac{1}{4},\frac{1}{4},\frac{1}{4},\frac{1}{4}%
,-1\right) \frac{\varepsilon }{\left( y+1\right) ^{2}}.  \label{S}
\end{equation}

We can view (\ref{5ds2}) as a five-dimensional analogue of the
Friedmann-Robertson-Walker cosmological metric for radiation, with an energy
density given by $\rho \left( y\right) =-\frac{\varepsilon }{\left(
y+1\right) ^{2}}$ (as measured by observers $\partial _{y}$).

Consider now a four-dimensional space $\left( M^{4},g\right) $ with a
vanishing Ricci tensor, i.e., a vacuum solution of the Einstein field
equations. Let us consider the question of embedding $\left( M^{4},g\right) $
into the collection ${\cal M}_{\tilde{g}_{0}}^{5}$ of five-dimensional
spaces that are Ricci-equivalent to $\left( \tilde{M}_{0}^{5},\tilde{g}%
_{0}\right) $. In order to work with a mixed Ricci tensor we redefine the
Ricci-equivalence property by the equation 
\begin{equation}
\tilde{R}_{\nu }^{\mu }=\frac{\partial \bar{f}^{\mu }}{\partial x^{\alpha }}%
\frac{\partial x^{\beta }}{\partial \bar{f}^{v}}S_{\beta }^{\alpha }.
\label{Rmisto}
\end{equation}

To find the embedding we begin with the ansatz 
\begin{eqnarray}
\bar{g}_{ik}\left( x^{1},...,x^{4},y\right) &=&u\left( y\right) g_{ik}\left(
x^{1},..,x^{4}\right) \\
\bar{\phi}\left( x^{1},...,x^{4},y\right) &=&1 \\
\bar{f}^{\alpha }\left( x^{1},...,x^{4},y\right) &=&x^{\alpha },
\end{eqnarray}
where $g_{ik}$ is the metric of $\left( M^{4},g\right) $ and $u(y)$ is a
function such that $u(0)=1.$

From Lemma 1 we can show that (\ref{Rmisto}) is equivalent to the ordinary
differential equation 
\begin{equation}
u^{\prime }=\frac{4}{5}\frac{u}{\left( y+1\right) }.  \label{equ}
\end{equation}

Therefore, after integrating (\ref{equ}) we conclude that $\left(
M^{4},g\right) $ has a local embedding in the space 
\begin{equation}
^{5}ds^{2}=\left( y+1\right) ^{\frac{4}{5}}\left( g_{ik}dx^{i}dx^{k}\right) +%
\frac{24}{25}\varepsilon dy^{2},  \label{5embed}
\end{equation}
whose Ricci tensor is the same as $S_{\nu }^{\mu }$, given by (\ref{S}).
Finally, it is worth mentioning that although the spaces (\ref{5ds2}) and (%
\ref{5embed}) are Ricci-equivalent they are not isometric. This can simply
be verified since the Weyl tensor $W_{\mu \nu \lambda \rho }$ calculated
from (\ref{5ds2}) vanishes while (\ref{5embed}) may have $W_{\mu \nu \lambda
\rho }\neq 0$ for some $g_{ik}$ (choose, for example, $g_{ik}dx^{i}dx^{k}$
to be the line element of Schwarzschild spacetime).

\section{Final comments}

The restriction of the Ricci-tensor being non-degenerate, as required by
Theorem 2, certainly imposes a limitation on the set of possible sources of
the embedding space. For example, we would have to leave out of
consideration solutions of the Einstein equations such as cosmological
models sourced by dust-type perfect fluid. We feel that although a great
number of solutions of physical interest have non-degenerate Ricci-tensor,
e.g. Friedman-Robertson-Walker models sourced by incoherent radiating
perfect fluids, it seems indisputable that a theorem in which the condition
of non-degeneracy is relaxed would be most welcome.

\section{Appendix}

{\bf Theorem (Cauchy-Kowalewski). }{\it Let us consider the set of partial
differential equations}$:$%
\begin{equation}
\frac{\partial ^{2}u^{A}}{\partial \left( y^{n+1}\right) ^{2}}=F^{A}\left(
y^{\alpha },u^{B},\frac{\partial u^{B}}{\partial y^{\alpha }},\frac{\partial
^{2}u^{B}}{\partial y^{\alpha }\partial y^{i}},\right) ,\qquad A=1,...,m
\label{CK}
\end{equation}
{\it where }$u^{1},..,u^{m}$ {\it are }$m$ {\it unknown functions of the }$%
n+1$ {\it variables }$y^{1},...,y^{n},y^{n+1},$ $\alpha =1,...,n+1,$ $%
i=1,..,n,$ $B=1,...,m.$ {\it \ Also, let }$v^{1},...,v^{m},w^{1},...,w^{m},$ 
{\it functions of the variables }$y^{1},...,y^{n},$ {\it be analytic at }$%
0\in ${\it \ }${\Bbb R}^{n}.${\it \ If the functions }$F^{A}$ {\it are
analytic with respect to each of their arguments around the values evaluated
at the point }$y^{1}=...=y^{n}=0,${\it \ then there exists a unique solution
of equations (\ref{CK}) which is analytic at }$0\in {\Bbb R}^{n+1}$ {\it and
that satisfies the initial condition} 
\begin{eqnarray}
u^{A}\left( y^{1},...,y^{n},0\right) &=&v^{A}\left( y^{1},...,y^{n}\right) \\
\frac{\partial u^{A}}{\partial y^{n+1}}\left( y^{1},...,y^{n},0\right)
&=&w^{A}\left( y^{1},...,y^{n}\right) ,\qquad A=1,...,m.
\end{eqnarray}

\section{Acknowledgments}

The authors thank Prof. M. Dajczer (IMPA) and J. Fonseca (UFPB) for helpful
discussions. This work was supported financially by CNPq.

\end{document}